\documentclass[preprint2]{aastex}
\usepackage{psfig}
\include{epsf}
\newcommand{\be}{\begin{equation}}
\newcommand{\ee}{\end{equation}}
\newcommand{\ba}{\begin{eqnarray}}
\newcommand{\ea}{\end{eqnarray}}

\def\simless{\mathbin{\lower 3pt\hbox
   {$\rlap{\raise 5pt\hbox{$\char'074$}}\mathchar"7218$}}}
\def\simgreat{\mathbin{\lower 3pt\hbox
   {$\rlap{\raise 5pt\hbox{$\char'076$}}\mathchar"7218$}}}   % > or of order

\shorttitle{SCUBA/Spitzer investigation of the submm background}
\shortauthors{Dye et al.}

\begin{document}

%% one line. However, use \\ to force a line break.

\title{An investigation of the submillimeter background radiation using
SCUBA and Spitzer}

%% from AASTeX v4.0. \email can be used to mark an email address
%% As in the title, \\ forces line breaks.

\author{S. Dye\altaffilmark{1}, S. A. Eales\altaffilmark{1},
M. L. N. Ashby\altaffilmark{2}, J. -S. Huang\altaffilmark{2},
T. M. A. Webb\altaffilmark{3}, P. Barmby\altaffilmark{2},
S. Lilly\altaffilmark{4}, 
M. Brodwin\altaffilmark{5},
H. McCracken\altaffilmark{6},
E. Egami\altaffilmark{7},
G. G. Fazio\altaffilmark{2}}
\affil{}
\altaffiltext{1}{School of Physics \& Astronomy, Cardiff University, 
5 The Parade, Cardiff, CF24 3YB, UK.}
\altaffiltext{2}{Harvard Smithsonian Centre for Astrophysics, 60 Garden Street,
Cambridge, MA 02138}
\altaffiltext{3}{Sterrewacht Leiden, Neilss Bohrweg 2, Leiden 233CA, 
Netherlands}
\altaffiltext{4}{Institute of Astronomy, Swiss Federal Institute of 
Technology Zurich, CH-8093 Zurich, Switzerland}
\altaffiltext{5}{JPL, CalTech, M/S 169-506, 4800 Oak Grove Drive, 
Pasadena, CA 91109, USA}
\altaffiltext{6}{Institute d'Astrophysique, 98bis, Bd Arago - 75014 Paris, 
France}
\altaffiltext{7}{Steward Observatory, University of Arizona, 933
North Cherry Avenue, Tuscon, AZ 85721}

\begin{abstract}

We investigate the redshift dependence of the contribution to the
extragalactic far-infrared/sub-millimeter background from galaxies
detected by the Spitzer Space Telescope at 8$\mu$m and 24$\mu$m. Using
seven-band optical to mid-infrared photometry, we estimate photometric
redshifts for the Spitzer sources which appear to be mostly L$_*$
galaxies at a median redshift of $z=1.0$.  These sources,
extracted from deep 8$\mu$m and 24$\mu$m mosaics of the CUDSS 14-hour
field with 5$\sigma$ limits of 5.8$\mu$Jy and 70$\mu$Jy respectively,
exhibit significant $850\mu$m and $450\mu$m emission as observed by
SCUBA. At $850\mu$m, after removing $\geq4\sigma$ sources and those
securely identified in our previous cross-matching paper, we measure
stacked flux at the significance level of $4.4\sigma$ and $2.9\sigma$
from the full $8\mu$m and $24\mu$m galaxy catalogue respectively. At
$450\mu$m, flux is detected from all $8\mu$m galaxies at the level of
$3.5\sigma$, while there is no significant emission from the $24\mu$m
galaxies.  We find that the 850$\mu$m flux is emitted almost
exclusively at $z \gtrsim 1.3$ from the Spitzer sources with 0.44mJy
($4.7\sigma$) per 8$\mu$m source and 0.51mJy ($2.8\sigma$) per
24$\mu$m source. This corresponds to a contribution of $(16\pm 3)\%$
toward the $850\mu$m extra-galactic background from the $8\mu$m
sources and $(5.0\pm 1.8)\%$ from the $24\mu$m sources.  At $450\mu$m,
only the 8$\mu$m sources within the redshift interval $1<z<2$ exhibit
significant emission with an average flux per source of 3.35mJy
($3.0\sigma$). This is a contribution of $(37\pm 12)\%$ to the
$450\mu$m background.

\end{abstract}

%% See the instructions to authors to determine appropriate 
%% keyword punctuation.


\section{Introduction}

Steady advances toward a thorough understanding of the population of
high redshift sub-millimeter (submm) sources uncovered by the
Sub-millimeter Common User Bolometric Array (SCUBA) and the Max-Planck
Millimeter Bolometer (MAMBO) have been made since their detection in
the first deep submm survey by \citet{smail97}. The two most
controversial issues concerning their nature have been, firstly,
identification of their dust-cloaked energy source and secondly, how
they relate to local systems.

The question regarding the energy source has now been largely
satisfied. The lack of strong X-ray emission from these sources
\citep[eg.][]{ivison00,fabian00,alexander03,almaini03,waskett03,alexander05}
suggests that their submm flux is dominated by re-radiated emission
from intense star formation, rather than AGN output.  Observations
constrain the net contribution of AGN activity to the level of $\sim
30\%$ \citep[see][and references contained
therein]{chapman05}. However, the question regarding their
relationship with the local galaxy population remains only partially
answered.

Finding the connection between distant submm sources and local
galaxies requires an understanding of their evolution.  Sources
detected in surveys with SCUBA and MAMBO have extremely high estimated
star formation rates (100 - 1000 $M_{\odot}$ yr$^{-1}$), enough to
rapidly form a massive elliptical
\citep{smail97,hughes98,lilly99,dunne03}.  However, it is not clear
from studies of the spatial density or clustering of submm sources
whether a direct link between local ellipticals and the submm
population can be made quite so confidently
\citep{fox02,scott02}.

The main observational limitation in understanding the SCUBA
population is the relatively large beam size. This gives rise to a
correspondingly large uncertainty on extracted source positions and
also means that confusion is a concern in deeper observations.  One of
the key ingredients needed to help resolve the evolutionary
relationship between local and distant submm galaxies, determination
of submm galaxy redshifts, is therefore especially difficult.  A
successful technique is to observe the sources using radio
interferometry, as the surface density of radio sources is
sufficiently low to make confident associations with coincident submm
sources. The high astrometric precision in these radio data greatly
eases the identification of optical counterparts for follow-up
spectroscopy.  Around two thirds of the submm sources detected at
850$\mu$m appear to have radio counterparts
\citep[eg.][]{ivison02,borys04}, but because the ratio of radio to
submm flux decreases with increasing redshift \citep{carilli99}, radio
observations are biased toward lower redshift objects
($z\lesssim3$). This method was recently applied by \citet{chapman05}
using Keck spectroscopy to determine the redshifts of 73 submm sources
cross-identified in VLA maps. They found a median redshift of
$z\sim2.2$.

A frequently ignored issue is that the sources in the bright SCUBA
samples may not be representative of the extragalactic far-infrared/submm
background radiation as a whole. Determining the nature of the
individual sources that constitute the background is of
the greatest importance because approximately 50\% of the entire
extragalactic background radiation (minus the CMB) is radiated in the
submillimeter waveband \citep{fixsen98}. Yet the bright SCUBA sources
for which there are redshifts may represent only a tiny fraction of
this background. The problem is that the far-IR/submillimeter background
is much brighter ($\sim 30$ times greater in terms of $\nu I_{\nu}$)
where it peaks at $\sim$200$\mu$m than it is at 850$\mu$m, the
wavelength of the SCUBA surveys.  At 850$\mu$m, around 30\% of the
background can be resolved into sources brighter than 3mJy
\citep[eg.][]{hughes98,eales00,smail02,webb03b,chapman05}. Estimates
of this fraction extend up to 60\% for sources down to 1mJy
\citep[eg.][]{smail02,chapman05}. The sources for which 
\citet{chapman05} measured redshifts are radio-detected sources with 
850$\mu$m fluxes brighter than $\simeq$3 mJy. Using estimates of the
spectral energy distributions of individual galaxies \citet{chapman05}
concluded that their sample represents 30\% of the background at
850$\mu$m but only about 2\% of the background at 200$\mu$m.
 
Therefore, despite the success of the work on the brighter SCUBA
samples, there are many unanswered questions. For example, is the
redshift distribution that \citet{chapman05} measure representative of
the far-IR/submm background as a whole? Until the background at
200$\mu$m can be directly resolved into individual sources, the next
best alternative is to study the SCUBA sources at fainter 850$\mu$m
fluxes. Unfortunately, here one hits the obstacle of confusion.

There are two ways round this obstacle. One is to exploit
gravitational lensing to mitigate the problem
\citep[eg.][]{smail02}. The second is the approach we adopt in this
paper, using the SCUBA images to estimate the submillimeter fluxes of
objects selected in other wavebands for which there are already
accurate positions. \citet{peacock00} used this reverse procedure by
stacking 850$\mu$m SCUBA detected emission from optical sources in the
Hubble Deep Field North. This study identified significant submm
emission from galaxies with high UV star formation
rates. \citet{serjeant03} repeated this work replacing the optical
data with 15$\mu$m ISO sources, but measured no emission. However, the
much improved sensitivity and resolution of the Spitzer Space
Telescope \citep{werner04} recently enabled \citet{serjeant04} to
statistically detect $5.8\mu$m and $8\mu$m sources in 850$\mu$m and
450$\mu$m SCUBA data. Finally, \citet{knudsen05} recently detected
significant 850$\mu$m emission from distant red galaxies selected by
J$-$K$>2.3$, finding that these sources are probably strong contributors
to the far-IR/submm background.

In this paper, we stack $450\mu$m and $850\mu$m flux observed by SCUBA
at the position of $8\mu$m and $24\mu$m Spitzer sources in the
Canada-United Kingdom Deep Sub-millimeter Survey (CUDSS) 14-hour
field.  From these data, we estimate the fractional contribution from
Spitzer sources to the extragalactic background at
850$\mu$m and 450$\mu$m.  Using a combination of ground-based optical,
near infrared and Spitzer 3.6$\mu$m and 4.5$\mu$m observations, we
compute photometric redshifts for the Spitzer sources, to investigate
the epoch at which their attributed submm flux is emitted.  This paper
follows on from our previous paper \citep{ashby05} where the forward
approach of cross-identifying Spitzer sources with submm sources was
carried out.

The layout of this paper is as follows: In Section \ref{sec_data} we
describe the data. Section \ref{sec_photo_z} details our photometric
redshift estimation. Stacking is carried out in Section
\ref{sec_stacking}. We conclude with a summary and discussion in
Section \ref{sec_discussion}.

\section{Data}
\label{sec_data}

Observations of the CUDSS 14 hour field analysed in this paper
comprise three distinct sets: 1) Mid-infrared observations using the
Spitzer Space Telescope, 2) Submm data observed with SCUBA, 3)
Ground-based optical and near infrared observations.

\subsection{Spitzer Space Telescope Data}

The Spitzer observations discussed in this paper were obtained as part
of the Guaranteed Time Observing program number 8 to image the
extended Groth strip, a $2^{\circ} \times 10'$ area at $\alpha \sim
14^{\rm h} 19^{\rm m}$, $\delta \sim 52^{\circ}48'$ (J2000) with the
Infra-red Array Camera \citep[IRAC;][]{fazio04} and Multi-band Imaging
Photometer for Spitzer \citep[MIPS;][]{rieke04}.  Approximately 90\%
of the CUDSS 14 hour field falls robustly inside this area, the
remaining 10\% in the south-east corner having poor or no coverage.
This corner was cropped in all data sets presented here (see Figure
\ref{scuba_map}).

Carrying on from our identification paper \citep{ashby05}, we have
used the 8$\mu$m IRAC and 24$\mu$m MIPS source positions for the
stacking.  The fraction of sources detected in the higher frequency
IRAC data that are dusty submm emitters is expected to be lower than
in the 8$\mu$m and 24$\mu$m data. In addition, the sensitivity of
these shorter wavelength data is much higher than the SCUBA maps so
that the stacking tends to sample noise rather than the detectable
population of faint sources. Indeed, \citet{serjeant04} failed to
detect significant submm emission from Spitzer sources observed in the
IRAC $3.6\mu$m and $4.5\mu$m channels in SCUBA observations of the
Hubble Deep Field. The converse is true of the longer wavelength MIPS
data where the number density of sources seen is too low to act as a
good probe of the submm emission observed in our SCUBA maps.  The
$5\sigma$ point source sensitivity of the $8\mu$m and 24$\mu$m data is
5.8$\mu$Jy and $70 \mu$Jy respectively.

Although we made no use of the $3.6\mu$m and $4.5\mu$m IRAC detected
sources in the stacking directly, photometry from these channels was
combined with the ground-based optical data for computation of
photometric redshifts in Section \ref{sec_photo_z}. The $5\sigma$
point source sensitivity of both the $3.6\mu$m and $4.5\mu$m data is
$0.9\mu$Jy.

There are 553 8$\mu$m sources that fall within the
850$\mu$m map area and 511 that fall within the 450$\mu$m map. Of the
24$\mu$m sources, 141 fall within the 850$\mu$m map and 133 within the
450$\mu$m map. All of the 24$\mu$m sources are detected
at 8$\mu$m.

\subsection{Sub-mm data}
\label{sec_submm_data}

The SCUBA observations amount to 63 hours worth of data taken on 20
different nights over the period from March 1998 to May 1999 at the
James Clark Maxwell Telescope (JCMT). The final 450$\mu$m and
850$\mu$m maps of the $\sim 7' \times 6'$ survey region were created
by combining many 'survey units'. Each survey unit was observed for
approximately one hour with a 64-point jiggle pattern (ensuring they
are fully sampled), nodding JCMT's secondary mirror and chopping by
$30''$ in right ascension.

The data reduction was carried out with the SURF package 
\citep{jenness97} and a noise map created using a Monte Carlo
technique to simulate the noise properties of each of SCUBA's
bolometers. We refer the reader to \citet{eales00} for more specific
details of the data reduction. 

\begin{figure*}[ht]
\epsfxsize=167mm
{\hfill
\epsfbox{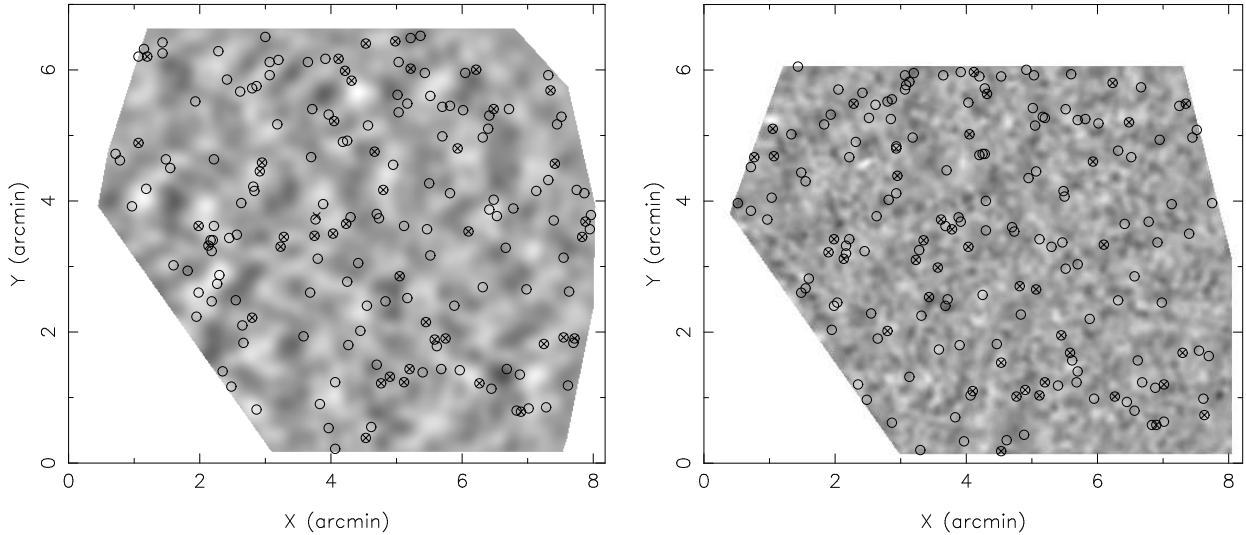}
\hfill}
\caption{The CUDSS 14 hour field
showing IRAC 8$\mu$m sources (circles) and MIPS 24$\mu$m sources
(crosses) overlaid on the SCUBA data. {\em Left:} The 850$\mu$m
beam-convolved map with $z>1.3$ Spitzer sources overlaid (lighter
shading indicates a higher submm flux). In this map, the `secure' and $\ge
4\sigma$ SCUBA sources have been removed (see text).  {\em Right:} The
450$\mu$m beam-convolved map with $1<z<2$ Spitzer sources overlaid. In
both maps, the south-eastern (bottom-left) corner is cropped where
there is no Spitzer coverage.}
\label{scuba_map}
\end{figure*}

We work with beam-convolved signal-to-noise (S/N) maps in this paper,
to maximise source detection. Since the goal is to detect faint SCUBA
sources buried in the map noise, prior to convolving with the beam, we
subtracted SCUBA sources based on their submm S/N and their
identifications in our previous paper \citep{ashby05}. For the
850$\mu$m data, we made three maps with different combinations of
SCUBA sources removed to assess whether the stacking is influenced by
this removal process. These three combinations are: 1) Removal of only
the sources listed in \citet{ashby05} with a secure Spitzer
identification.  A `secure' source is defined as having a bright
$8\mu$m positionally coincident ($r<10''$) Spitzer counterpart with
IRAC-MIPS colors indicative of a high-redshift dusty source
\citep[see][]{huang04}. \citet{ashby05} made 7 secure identifications.
2) Removal of the secure sources along with the remaining 5 sources
detected by SCUBA with significance $\ge 4\sigma$. 3) Removal of the
12 sources in case 2 along with the remaining 6 sources having a
`possible' Spitzer identification at 8$\mu$m. A source identification
is classified as `possible' when selected from multiple co-incident
sources. As in case 1, the most likely counterpart is chosen based on
IRAC-MIPS colors.  Removing this last combination of 18 sources leaves
only 2 of the 23 SCUBA sources in the 850$\mu$m data (three fall in
the cropped south-eastern corner), both with significances of
$3.0\sigma$ \citep{webb03}.

Sources were removed from the raw SCUBA maps by subtracting the beam
profile at the position of the source, scaled by the source
flux. There are no directly detectable significant sources in the
450$\mu$m data, hence no sources were removed from the 450$\mu$m map.
Figure \ref{scuba_map} shows the beam-convolved S/N map at 850$\mu$m
with the secure + $\ge 4\sigma$ sources removed and the (unmodified)
map at 450$\mu$m.  The 8$\mu$m and 24$\mu$m Spitzer source positions
within the CUDSS 14h area having $z>1.3$ are over-plotted on the
850$\mu$m map and those with $1<z<2$ are over-plotted on the 450$\mu$m
map, to coincide with the findings of Section \ref{sec_z_analysis}.

\subsection{Optical \& Near Infrared Data}

In addition to the IRAC matched photometry, we also matched to optical
and near-infrared ground-based data. For the optical photometry, we
extracted U, B, V \& I fluxes from the Canada-France Deep Fields survey
(CFDF) \citep{mccracken01} imaged with the 4.0m Blanco Telescope and
the Canada-France-Hawaii Telescope (CFHT). Like the IRAC photometry,
fluxes were summed in $3''$ diameter apertures. The $3\sigma$ point
source sensitivities in AB mags are 27.71, 26.23, 25.98 and 25.16 for
U, B, V \& I respectively.

For the near-infrared photometry, we matched to our own SExtracted
\citep{bertin96} catalogue of the K-band image of \citet{webb03},
observed with the CFHT using the CFHTIR camera. The K-band data
reach a depth of $K_{\rm AB}\sim 23$.

\section{Photometric Redshifts}
\label{sec_photo_z}

From the matched aperture photometry in U, B, V, I \& K as well as the
IRAC 3.6$\mu$m and 4.5$\mu$m channels, we obtained photometric
redshifts for the 8$\mu$m and 24$\mu$m sources using the HyperZ
redshift code \citep{bolzonella00}. We decided not to use photometry
from the two longest-wavelength IRAC channels, 5.8$\mu$m \& 8$\mu$m,
because these data are less sensitive than the two shortest-wavelength
channels, they are more affected by dust emission features
\citep{lu03} and the spectral templates are uncertain at these
wavelengths (specifically, the contribution from
PAHs).  We excluded sources with poor quality measurements at
3.6$\mu$m and 4.5$\mu$m, although we did not exclude sources with poor
measurements or non-detections in the longer wavelength bands.

The local galaxy spectral energy distributions 
(SEDs) packaged with HyperZ are those given by
\citet{coleman80}, with an extrapolation into the infrared using the
results of spectral synthesis models.  The lack of any empirical basis
for these templates at wavelengths $>1\mu$m is clearly unsatisfactory
since our photometry includes both near-infrared and mid-infrared
measurements. Therefore, we constructed our own set of templates in
the following way. \citet{mannucci01} list empirical SEDs extended
from 0.1$\mu$m to 2.4$\mu$m for the Hubble types E, S0, Sa, Sb
and Sc. We extended these out to 6$\mu$m using the average SEDs for
disk galaxies and elliptical galaxies listed in Table 3 of \citet{lu03}
using the disk-galaxy SED for all Hubble types apart from ellipticals.
The one disadvantage of these templates is the lack of a
template for an irregular galaxy, and we therefore retained the HyperZ
template for an Im galaxy, extending this into the mid-infrared using
the average disk-galaxy SED from \citet{lu03}.

\begin{figure}[ht]
\epsfxsize=75mm
{\hfill
\epsfbox{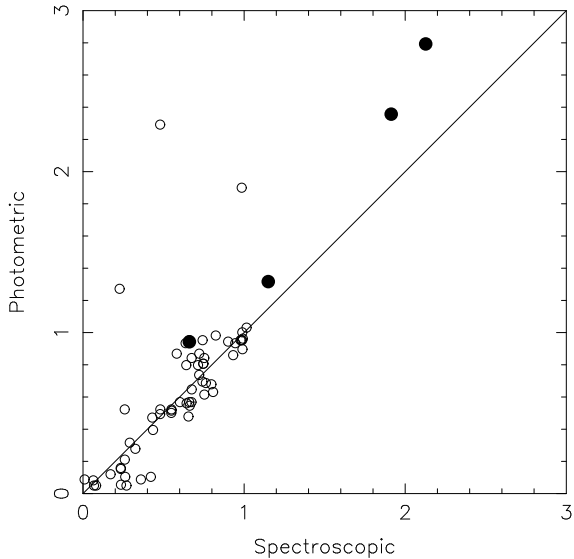}
\hfill}
\caption{Comparison of photometric redshifts (using U, B, V, I, K, 
3.6$\mu$m \& 4.5$\mu$m fluxes) with spectroscopic redshifts for the
8$\mu$m sources in the CUDSS 14h field. Open circles show the 61
8$\mu$m sources with redshifts measured by the CFRS. Filled
circles are the 4 SCUBA sources with spectroscopic redshifts measured
by \citet{chapman05}.}
\label{photo_vs_spec_z}
\end{figure}

We tested the accuracy of our photometric redshifts by estimating
redshifts for the 8$\mu$m sources in the field that have
spectroscopically confirmed redshifts. There are 61 galaxies from the
Canada-France Redshift Survey (CFRS) \citep{lilly95} that are
detected by Spitzer at 8$\mu$m and an
additional four objects that are possible SCUBA detections with
spectroscopic redshifts \citep{chapman05}. Figure
\ref{photo_vs_spec_z} shows the photometric redshift estimates plotted
against the spectroscopic redshifts.  The agreement is fairly good
with the scatter, defined as $\sqrt{\Sigma [\Delta z / (1 + z)]^2/N}$,
equal to 0.217 for the CFRS sources and 0.161 for the four SCUBA
galaxies. We note, however, that there are very few galaxies with
spectroscopic redshifts at $z > 1$. We are therefore less confident
about the accuracy of estimates beyond this redshift, although the
large bin size used when stacking by redshift in Section
\ref{sec_z_analysis} will greatly reduce the impact of large errors.

\begin{figure}[ht]
\epsfxsize=70mm
{\hfill
\epsfbox{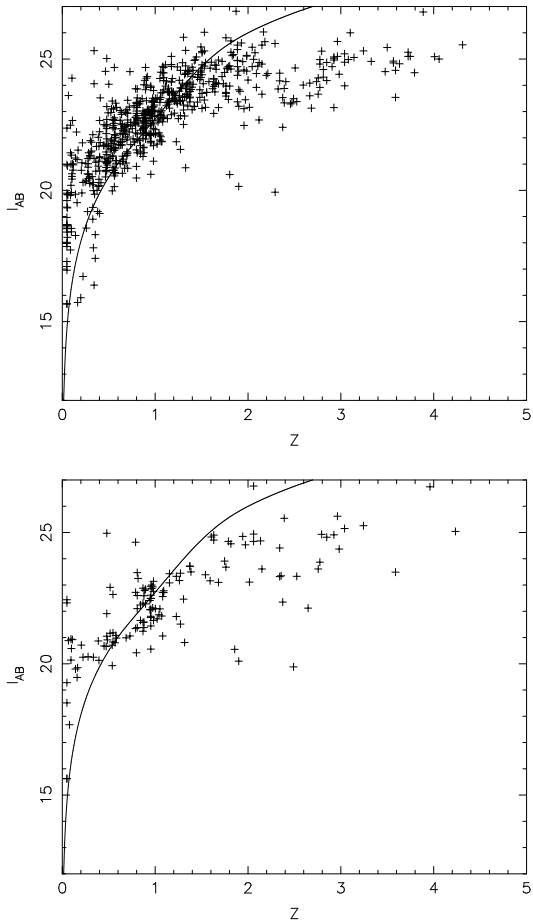}
\hfill}
\caption{I-band magnitude versus photometric redshift for all 8$\mu$m
sources ({\em top}) and 24$\mu$m sources ({\em bottom}). Solid line
shows prediction for a non-evolving L$_*$ Sbc galaxy.}
\label{I_vs_photo_z}
\end{figure}

Figure \ref{I_vs_photo_z} shows the I-band magnitude versus
photometric redshift for all the objects in our 8$\mu$m and 24$\mu$m
catalogue. The curve shows the prediction for a non-evolving L$_*$
galaxy, with the K-correction calculated using the SED for an Sbc
galaxy.  We have used the value for L$_*$ given by
\citet{blanton01}. The diagram shows that the galaxies detected by
Spitzer at 8$\mu$m are mostly L$_*$ galaxies with a fairly small
dispersion about this luminosity.  This small dispersion is additional
confirmation that our photometric-redshift technique works
reasonably well (approximately 50\% more scatter is seen if the
Spitzer photometry is omitted in the redshift analysis).

\section{Stacking analysis}
\label{sec_stacking}

The stacking process determines the sum of the submm flux detected at
the position of Spitzer sources and its significance. In this work, 
we calculate the total weighted flux, given by
\be
\label{eq_av_wtd_flux}
f=N(\Sigma_i f_i \sigma_i^{-2})/ (\Sigma_i \sigma_i^{-2}),
\ee
where $f_i$ is the submm flux taken from the beam-convolved map
at the position of the $i$th Spitzer source,  $\sigma_i$ is its error
and $N$ is the total number of sources stacked.

We measure the significance of the stacking in two separate ways:

\begin{itemize}

\item[-] The first uses the Kolmogorov-Smirnov (KS) test to quantify the
significance of the discrepancy between the distribution of S/N drawn
from the submm map at the position of the Spitzer sources and the S/N
distribution of all pixels in the entire map. If the former distribution is
positively skewed with respect to the latter, then this is attributed
to the detection of faint SCUBA sources. 

\item[-] The second is based on the significance of the actual total
weighted flux measured. The formal error on $f$ is defined as
\be
\label{eq_sigma_f}
\sigma_f = N/\sqrt{\Sigma_i \sigma_i^{-2}}.
\ee
One can then define the significance $\beta=f/\sigma_f$.
However, \citet{serjeant03} showed using Monte Carlo simulations that
$\beta$ is not normally distributed. We correct for this with our own
Monte Carlo simulations (see below).

\end{itemize}

We determined the distribution of $\beta$ and the KS significance
by carrying out Monte Carlo simulations
for each of the four SCUBA datasets considered in this paper (the
three 850$\mu$m maps and the 450$\mu$m map). 100,000 
realisations of the Spitzer source positions were performed per
dataset. We found that the distribution of KS significance is
perfectly Gaussian in every case, whereas the distribution of
$\beta$ is skewed positively by an amount depending on the submm
dataset.

To account for the non-Gaussianity in $\beta$, we used its
distribution function calculated from the Monte Carlo analysis to make
plots showing the relationship between $\beta$ and its equivalent
Gaussian significance, $\beta_g$. All four SCUBA datasets gave linear
relationships which were fitted as follows: 850$\mu$m secure
$\beta_g=0.79\beta-0.15$; 850$\mu$m $\geq4\sigma$ + secure
$\beta_g=0.95\beta-0.09$; 850$\mu$m possible + secure
$\beta_g=1.01\beta-0.22$; 450$\mu$m $\beta_g=0.86\beta+0.26$. From
these relationships, it is apparent that the 850$\mu$m data with only
secure sources removed would quite heavily over-predict the stacking
significance if not corrected. The corrections for the remaining SCUBA
data are fairly modest.

In this paper, all total weighted flux significances quoted are
the corrected version $\beta_g$. Absolute errors quoted
on the total weighted flux are those given by equation (\ref{eq_sigma_f}).

Despite the fact that the KS significance adheres to Gaussian
statistics, we found that it often gives false peaks in simulations of
alignment maps described in the following section, whereas the
corrected weighted flux significance, $\beta_g$, does not.  For this
reason, we use $\beta_g$ to assess data alignment.  However, in
Section \ref{sec_z_analysis} where we discuss stacking Spitzer sources
selected by redshift, we calculate both $\beta_g$ and the KS
significance for comparison.

\subsection{Stacking the full 8$\mu$m \& 24$\mu$m catalogues}
\label{sec_stack_no_z}

A crucial assumption in the stacking analysis is that the astrometric
solutions of the Spitzer and SCUBA datasets are not offset from each
other. We therefore tested the alignment of the SCUBA and Spitzer data
by calculating $\beta_g$ for a range of offsets in RA and Dec added
to the Spitzer source positions. The test plots
this as a map with the x-axis corresponding to the RA offset and the
y-axis the Dec offset. If the Spitzer data is well aligned with the
SCUBA maps {\em and} if a significant detection is made, a peak in the
vicinity of the origin should be apparent. This peak should be
singular. If more than one significant peak is seen, then the
alignment of both data sets cannot be established with confidence and
the robustness of the detection must also be drawn into question.

\begin{figure*}[ht]
\epsfxsize=110mm
{\hfill
\epsfbox{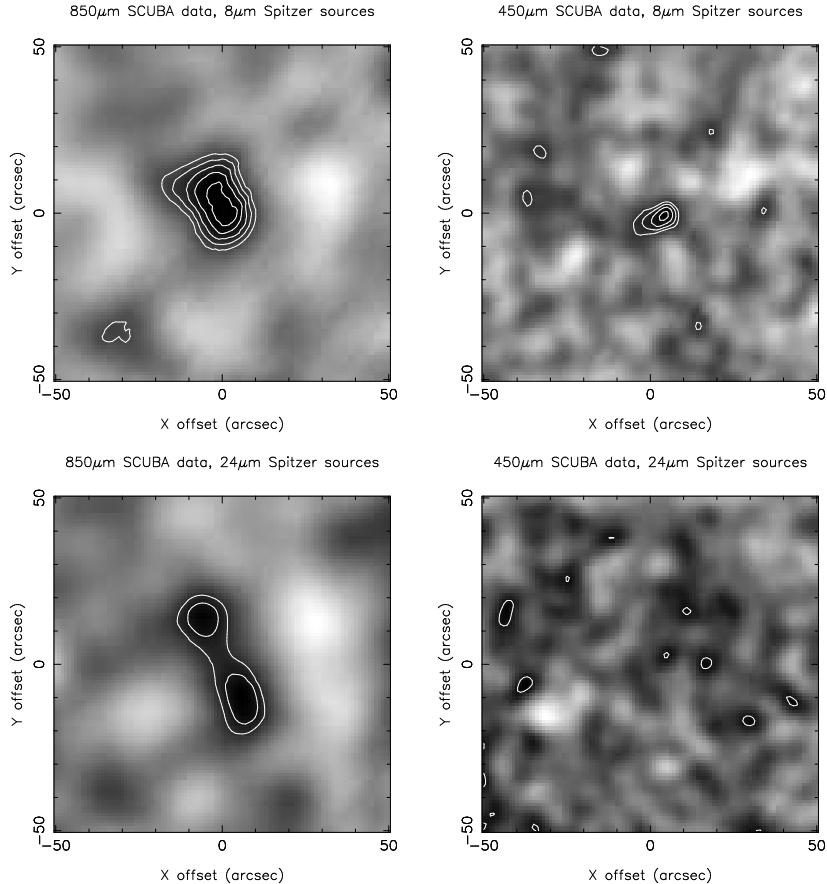}
\hfill}
\caption{Variation of the total weighted submm flux significance,
$\beta_g$, as a function of the offset of the SCUBA maps from the
Spitzer source catalogue (i.e. add offset to Spitzer position to get
SCUBA position).  Contours start at $2\sigma$ level and have intervals
of $0.5\sigma$.  Top row shows results for the 8$\mu$m Spitzer sources
and the bottom row shows the 24$\mu$m source results. The $850\mu$m
map has secure + $\ge 4 \sigma$ submm sources removed; no submm 
sources are removed from the $450\mu$m map.}
\label{corr}
\end{figure*}

Figure \ref{corr} shows the maps of $\beta_g$ over the $100''$ range
of offsets in RA and Dec spanned, for both 8$\mu$m and 24$\mu$m
Spitzer sources stacked with the 450$\mu$m data and the 850$\mu$m data
with secure + $\ge 4 \sigma$ sources removed.  The contours start at a
significance of $2 \sigma$ with intervals of $0.5 \sigma$. The more
smoothly varying features of the 850$\mu$m plots relative to the
450$\mu$m plots reflects the larger beam size at 850$\mu$m.

Stacking the 8$\mu$m sources shows a clear detection at both
850$\mu$m and 450$\mu$m with peak significances of 4.4$\sigma$ and
3.5$\sigma$ respectively. The 850$\mu$m peak occurs at an
offset of $(0'',-2'')$ and the 450$\mu$m peak at $(4'',0'')$.
The results from stacking with the 24$\mu$m
catalogue are less clear; at 850$\mu$m, there are two $2.9\sigma$
peaks at offsets of $(5'',-10'')$ and $(-8'',12'')$ with a
significance of $2.5\sigma$ at $(0'',0'')$. There is no
detection of the 24$\mu$m sources at 450$\mu$m.

%\FloatBarrier

To determine the significance of these offsets, we carried out Monte
Carlo simulations of the stacking at 850$\mu$m and 450$\mu$m. For each
realisation, we created a noisy synthetic SCUBA map and placed sources
with fixed S/N by adding the SCUBA beam at random positions. After
convolving the map with the beam, stacking was carried out, offsetting
this randomized catalogue by varying amounts to produce a $\beta_g$
map, like those shown in Figure \ref{corr}. We carried out 200
realisations for each of five different source S/N values.

\begin{figure}[ht]
\epsfxsize=70mm
\vspace{2mm}
{\hfill
\epsfbox{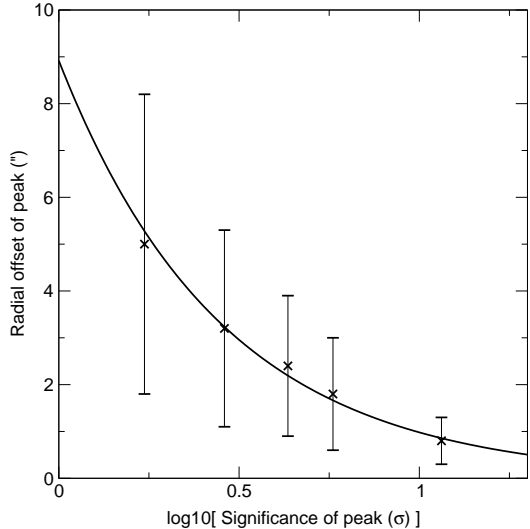}
\hfill}
\caption{Results of Monte Carlo simulations to determine the variation
of the mean and scatter of peak offset from $(0'',0'')$ in the
significance maps as a function of the amplitude of the peak
measured. Results plotted here apply to the $8\mu$m sources and either
850$\mu$m or 450$\mu$m data.}
\label{sims}
\end{figure}

Figure \ref{sims} shows the mean and scatter of the radial offset of
the peak in $\beta_g$ from $(0'',0'')$ as a function of the
significance of the peak. We found no discernible difference between
the 850$\mu$m and 450$\mu$m results hence this plot serves for both
wavelengths. From the plot, we conclude that the peak of $4.4\sigma$
seen in the significance map for the 8$\mu$m sources and 850$\mu$m
data, occurs on average in the simulations at a radial offset of
$r=2.0'' \pm1.5''$. The measured offset of $(0'',-2'')$ is therefore
consistent with there actually being no misalignment between the
850$\mu$m SCUBA map and 8$\mu$m data. Similarly, at $450\mu$m, the
offset of $(4'',0)$ of the $3.5\sigma$ peak is consistent with the
expected offset of $r=2.6''\pm1.7''$ for no misalignment.  The
implication that the 450$\mu$m and 850$\mu$m maps are well-aligned is
reassuring; as an additional check, we verified that this is the case
from astrometry of point source calibrators acquired at both
wavelengths throughout the CUDSS observing. The 24$\mu$m catalogue has
the same astrometry as the 8$\mu$m catalogue since both are tied to
the Two Micron All Sky Survey \citep[2MASS;][]{cutri03}, hence we
conclude that all datasets used in the stacking are properly aligned.

To summarise this section, we can make two statements. Firstly, there
is significant emission at 850$\mu$m and 450$\mu$m from all 8$\mu$m
Spitzer sources on average. From the 24$\mu$m sources, the total
weighted flux has a significance of $2.5\sigma$ at 850$\mu$m but
because the significance map is dual-peaked, this is a tentative
detection. Secondly, the Spitzer data and the SCUBA data are
well-aligned.

In the next section, we split the Spitzer catalogues by redshift to
investigate the possibility of this detection being weakened by
foreground/background Spitzer sources not truly associated with the
observed submm emission.

\subsection{Stacking by redshift}
\label{sec_z_analysis}

Having assigned photometric redshifts to the Spitzer catalogues, we
can stack sources separated into redshift bins. The bin size must be
large compared to the typical redshift uncertainty.  In addition, the
number of objects per bin must be equal so that the significance
across bins can be fairly compared (i.e. to keep Poisson noise
constant per bin).  Given the accuracy of our photometric redshifts
discussed in Section \ref{sec_photo_z}, we split the 8$\mu$m sources
into 6 redshift bins and the 24$\mu$m sources into 5. This gives an
average bin width of $\Delta z \sim 0.5$.  Stacking was then carried
out for each bin at both 850$\mu$m and 450$\mu$m.

\begin{figure*}[ht]
\epsfxsize=150mm
{\hfill
\epsfbox{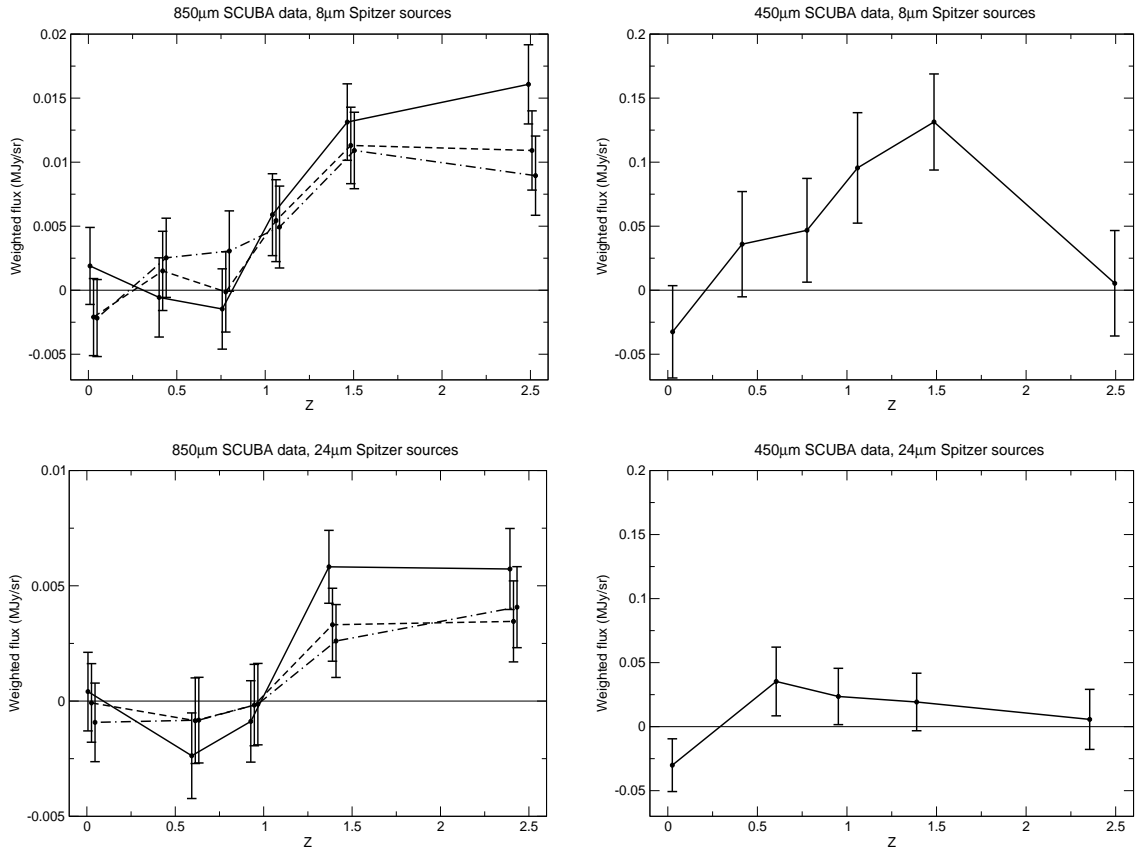}
\hfill}
\caption{Variation of total weighted flux of objects binned by redshift
(median bin redshift plotted). {\em Top left}: The 850$\mu$m data
stacked with the 8$\mu$m sources, where the solid, dashed and
dot-dashed lines indicate secure, secure + $\ge4 \sigma$ and secure +
possible SCUBA source subtraction, respectively. Abscissae are plotted
slightly staggered for clarity. {\em Top right}: The 450$\mu$m data
stacked with the 8$\mu$m sources. {\em Bottom left}: 850$\mu$m data
stacked with the 24$\mu$m sources.  {\em Bottom right}: 450$\mu$m data
stacked with the 24$\mu$m sources.}
\label{flux_vs_z}
\end{figure*}

The total weighted flux from all sources in a given redshift bin
(expressed in units of MJy/sr) is plotted against the median bin
redshift in Figure \ref{flux_vs_z}. The 850$\mu$m data show a clear
detection with both 8$\mu$m and 24$\mu$m Spitzer sources lying at $z
\gtrsim 1.3$.  With the 450$\mu$m data, only the 8$\mu$m Spitzer
sources produce a notable detection at $z
\sim 1.5$.

\begin{table*}
\centering
\small
\begin{tabular}{|l|c|c|c|c|}
\hline
SCUBA data & \multicolumn{2}{|c|}{KS Significance} & 
\multicolumn{2}{|c|}{Submm flux (MJy/sr)} \\ \cline{2-5}
 & 8$\mu$m sources & 24$\mu$m sources  & 8$\mu$m sources & 
    24$\mu$m sources\\

\hline

850$\mu$m $-$ secure   & 4.8$\sigma$ & 2.6$\sigma$ & 0.032$\pm$0.004 (172, $5.5\sigma$) & 0.011$\pm$0.002 (45, $3.9\sigma$) \\
850$\mu$m $-$ (secure + $\geq4\sigma$) & 4.7$\sigma$ & 3.4$\sigma$ & 0.024$\pm$0.004 (172, $5.1\sigma$) & 0.007$\pm$0.002 (45, $2.7\sigma$) \\
850$\mu$m $-$ (secure + poss.) & 4.5$\sigma$ & 3.3$\sigma$ & 0.022$\pm$0.004 (172, $4.7\sigma$) & 0.007$\pm$0.002 (45, $2.8\sigma$) \\
450$\mu$m   & 2.3$\sigma$ & $<1\sigma$  & 0.175$\pm$0.056 (170, $3.0\sigma$) &  0.035$\pm$0.027 (41, $1.4\sigma$) \\

\hline
\end{tabular}
\normalsize
\caption{Spitzer 8$\mu$m and 24$\mu$m source detections at 850$\mu$m and
450$\mu$m. Redshift selection $z>1.3$ applies to the 850$\mu$m data
and $1<z<2$ applies to the 450$\mu$m data. Listed are the
significances of the KS tests and the total weighted submm flux of all
objects within the corresponding redshift selection. Quantities in
brackets list the number of Spitzer galaxies stacked and the
significance of the weighted flux, $\beta_g$ (note that this is not
the ratio flux/error due to its non-Gaussian distribution -- see
text).}
\label{tab_fluxes}
\end{table*}

Choosing our most conservative 850$\mu$m dataset, that with all secure
and possible sources removed, and selecting Spitzer sources with $z >
1.3$, we found that the 8$\mu$m sources are detected at the level of
$4.7\sigma$ and the 24$\mu$m sources at $2.8\sigma$.  At 450$\mu$m,
the $8\mu$m sources selected by $1 < z < 2$ are detected with a
significance of $3.0\sigma$. The 24$\mu$m Spitzer sources selected
by $1 < z < 2$ are not detected.

\begin{figure*}[ht]
\epsfxsize=150mm
{\hfill
\epsfbox{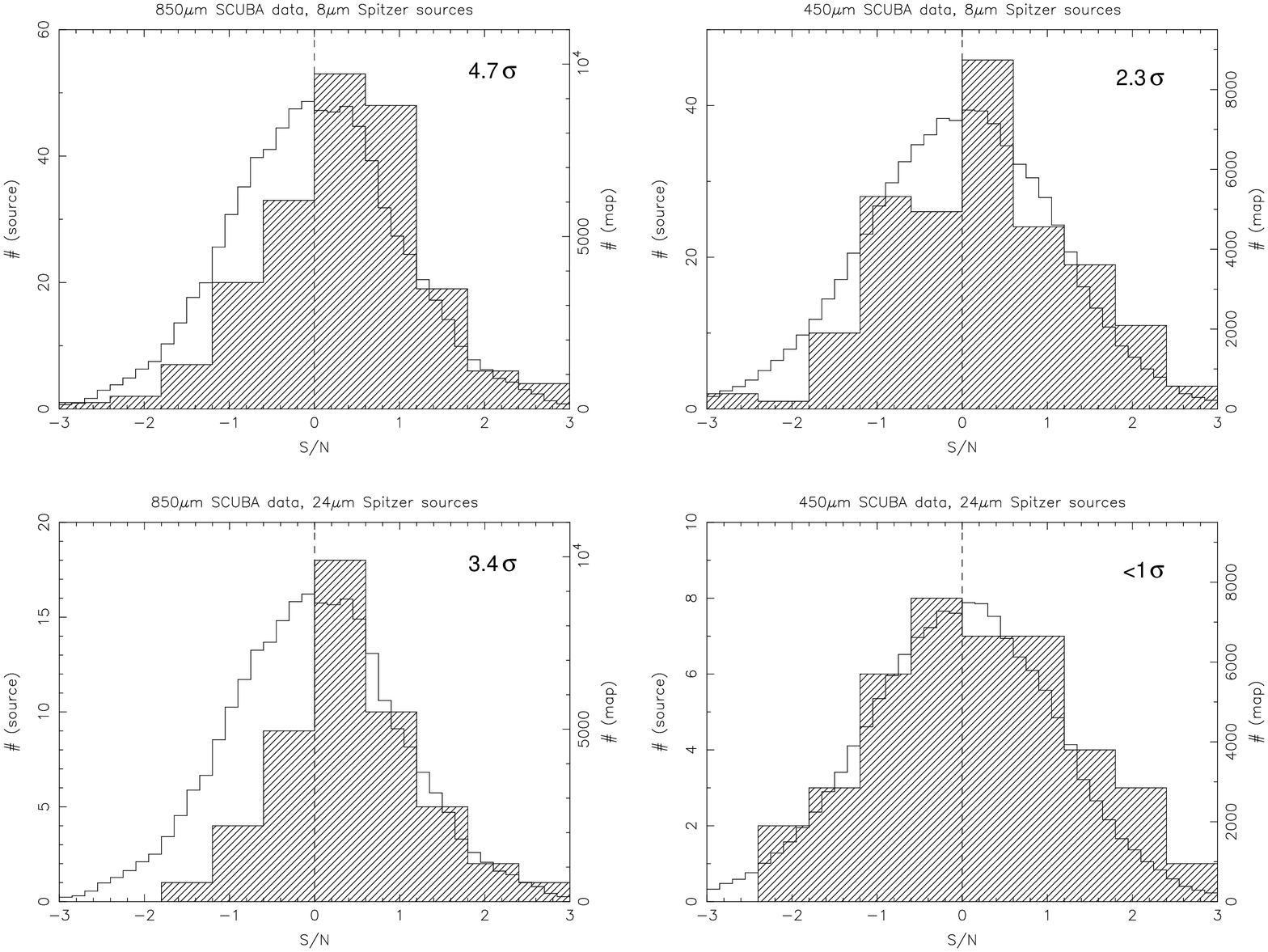}
\hfill}
\caption{Distribution of submillimeter S/N at the positions of the
Spitzer sources ({\em shaded histogram, left ordinates}) compared with
the distribution of S/N over the entire SCUBA map ({\em open
histogram, right ordinates}). For the 850$\mu$m data,
secure + $\ge 4\sigma$ SCUBA sources are removed and 
$z>1.3$ Spitzer sources are selected. No SCUBA sources
are removed from the 450$\mu$m data and $1<z<2$ Spitzer sources
are selected.}
\label{histograms}
\end{figure*}

These detections were verified with the KS test.  We applied the same
Spitzer source redshift selections and chose the 850$\mu$m data
with secure and possible sources removed. 
Figure \ref{histograms} shows the results. In these plots, the S/N
distribution of the full SCUBA map (the `map
distribution') is shown as an open histogram and the distribution of
S/N at the redshift selected Spitzer source positions
(the `source distribution') as a shaded histogram.  In
all plots, bar that for the 450$\mu$m $+$ 24$\mu$m data, excess
emission from faint submm sources causes the source histogram to be
clearly skewed positively with respect to the map histogram. The KS
test quantifies the significances of these skews as follows: 850$\mu$m
emission from the 8$\mu$m sources at $4.5\sigma$, 850$\mu$m emission
from the 24$\mu$m sources at $3.3\sigma$ and 450$\mu$m emission from
the 8$\mu$m at $2.3\sigma$. The 24$\mu$m sources are not detected at
450$\mu$m.

Table \ref{tab_fluxes} summarises the results for all versions of the
submm data. The weighted flux significances agree fairly well with the
KS significances, although on the whole, they are slightly higher.
The largest discrepancy occurs in the 850$\mu$m map with only secure
sources removed.  In this case, the $\geq 4
\sigma$ submm sources and/or the submm identifications defined as
possible by \cite{ashby05} are detected in the stacking.

\section{Summary and Discussion}
\label{sec_discussion}

We have measured significant 850$\mu$m emission from Spitzer 8$\mu$m
and 24$\mu$m sources and 450$\mu$m emission from the 8$\mu$m
sources. The $5\sigma$ point source sensitivities of the Spitzer
data are 5.8$\mu$Jy at 8$\mu$m and 70$\mu$Jy at 24$\mu$m. By
computing photometric redshifts for the Spitzer sources using optical,
near infra-red and Spitzer $3.6\mu$m and $4.5\mu$m photometry, we have
been able to statistically identify the epoch from which this submm
emission dominates. We find that the 850$\mu$m flux is almost
exclusively emitted from Spitzer sources at redshifts $z\gtrsim 1.3$
up to the highest redshift of $z \sim 4$ we measure in our sample. Our
most conservative estimate of this flux (i.e. having subtracted secure
and possible SCUBA sources) is 0.44mJy per 8$\mu$m Spitzer source
($4.7\sigma$ significance) and 0.51mJy per 24$\mu$m Spitzer source
($2.8\sigma$ significance).  At 450$\mu$m, the emission appears to
peak around $z\sim 1.5$ with an average stacked flux per 8$\mu$m
Spitzer source within $1<z<2$ of 3.35mJy ($3.0\sigma$ significance).

We can estimate the contribution our stacked flux makes to the
extragalactic far-IR/submm background using the spectrum of
\citet{fixsen98}.  According to their measurements, the background
flux at 850$\mu$m is 0.14MJy/sr.  The total weighted flux we measure
from $8\mu$m sources at $z>1.3$ in our 850$\mu$m map with secure and
possible sources removed is $0.022\pm0.004$MJy/sr, a contribution of
$(16\pm3)\%$. The 24$\mu$m sources at $z>1.3$  are responsible for
approximately one third of this with a fractional contribution of
$(5.0\pm1.8)\%$. At 450$\mu$m, \citet{fixsen98} measure a background
flux of 0.47MJy/sr. The weighted total flux from the 8$\mu$m
sources with $1<z<2$ in our 450$\mu$m data is $0.175\pm0.058$MJy/sr, some
$(37\pm12)\%$ of the background.

Our estimate of the 850$\mu$m contribution to the background from
8$\mu$m sources is consistent with the fraction of $(20\pm8)\%$
determined by \citet{serjeant04}. Within their errors,
\citet{serjeant04} find that the 450$\mu$m flux from 8$\mu$m sources
is sufficient to account for all of the background radiation at this
wavelength. This is in contrast to our findings that indicate a
maximum contribution of approximately 50\%.

Including the flux of submm sources removed from the 850$\mu$m data
gives a measure of the combined contribution of directly detected and
stacked sources to the background at this wavelength. The secure and
possible submm sources have a total weighted 850$\mu$m flux of
$(62\pm5)$mJy.  Adding this to the stacked 850$\mu$m flux from the
8$\mu$m sources gives a total fractional background contribution of
$(29\pm3)\%$. \citet{webb03c} estimate that Lyman break galaxies
(LBGs) contribute $\sim (20\pm10)\%$ of the 850$\mu$m background.
Even making the extreme assumption that there is absolutely no overlap
between the LBG and Spitzer 8$\mu$m population \citep[but see]
[who estimate an overlap of $\sim 25\%$]{huang05} 
the total fraction of the 850$\mu$m 
background that remains unaccounted for is at least $40\%$.

The average submm fluxes we measure in the stacking correspond to
fairly regular systems. To demonstrate this, we can use the definition
of an ultra-luminous infrared galaxy (ULIRG) set by
\citet{clements99}. This stipulates that a ULIRG must have a
luminosity of at least $10^{11.4}$L$_{\odot}$ measured at 60$\mu$m by
the infra-red astronomical satellite (IRAS). With an assumed IRAS
galaxy SED, the expected 850$\mu$m and 450$\mu$m flux can be
calculated for a ULIRG at the median redshift of our catalogue,
$z=1.0$.  Following \citet{clements04}, we take the coolest and
warmest IRAS galaxy SEDs from the sample of \citet{dunne01} to
estimate the range of submm emission expected. Using the SED of
NGC958, which is dominated by cold dust at 20K, the expected submm
flux from an object with a 60$\mu$m IRAS flux of
$10^{11.4}$L$_{\odot}$ at $z=1.0$ is 0.77mJy at 850$\mu$m and 2.06mJy
at 450$\mu$m (h$_{100}=0.7$, $\Omega_{\rm m}=0.3$, $\Lambda=0.7$). 
Similarly, the SED of IR1525$+$36 has a much warmer mix
of dust with temperatures 26K and 57K in the ratio 15:1, and predicts
a flux of 0.06mJy at 850$\mu$m and 0.20mJy at 450$\mu$m.  The
average stacked 850$\mu$m and 450$\mu$m fluxes of 0.44mJy and 3.35mJy
we measure from the 8$\mu$m Spitzer sources indicates that our
stacking is sensitive to borderline ULIRGS.  Although
ULIRGS would be considered extreme systems in the local universe, at
$z\sim 1$, they are quite the norm \citep[eg.][]{daddi05}.

An interesting result is the peak in 450$\mu$m emission we measure
from the 8$\mu$m sources around $z=1.5$. At this wavelength, the
energy density of the background is approximately 25\% of the maximum
at $\sim 200\mu$m (compared to only 3\% at 850$\mu$m). Given that the
8$\mu$m sources make up a significant fraction of the 450$\mu$m
background, it would not be unreasonable to expect that the entire
far-IR/submm background is dominated by systems around this redshift.
This is a slightly lower redshift than the median redshift $z=2.2$ of
the population of radio detected submm galaxies studied by
\citet{chapman05}. However, the Chapman et al. sample is substantially
brighter (by $\times 40$) on average than the submm galaxies probed by
the stacking in our work. This is consistent with the downsizing
scenario in which larger galaxies generally form earlier than smaller
galaxies. However, this result could be heavily influenced by
difficult to quantify, varying selection effects between both
studies. Clearly, we must wait for observations with a
high $200\mu$m sensitivity to provide a complete and definitive answer
as to the nature of the far-IR/submm background.

\vspace{5mm}
\begin{flushleft}
{\bf Acknowledgements}
\end{flushleft}

This work is based on observations made with the Spitzer Space
Telescope, which is operated by the Jet Propulsion Laboratory,
California Institute of Technology under NASA contract 1407. Support
for this work was provided by NASA through contract 1256790 issued by
JPL/Caltech.

\end{document}